\documentclass[useAMS,usenatbib]{mn2e}

\usepackage{aas_macros}
\usepackage{amsmath,amssymb,wasysym}
\usepackage[english]{babel}
\usepackage{ifpdf}
\usepackage[latin9]{inputenc}
\usepackage{mathpazo,mathrsfs,float,psfrag}
\usepackage[usenames,dvipsnames]{color}

\usepackage{hyperref}
\hypersetup{
  pdfauthor={Korobkin et al.},
  pdftitle={The runaway instability in general relativistic accretion disks},
  pdfsubject={Relativitic astrophysics},
  urlcolor=blue,
}

\usepackage[pdftex]{graphicx}
\usepackage{epstopdf}

\def\cm{\textrm{cm}}

\def\gr{\textrm{g}}
\def\sec{\textrm{s}}

\def\MSun{\mathcal M_{\astrosun}}
\def\Mbh{\mathcal M_{\rm BH}}
\def\Mdisk{\mathcal M_{\rm disk}}

\def\eqref#1{(\ref{#1})}

\def\Quilt{\textsc{Quilt}}
\def\Cactus{\textsc{Cactus}}
\def\Carpet{\textsc{Carpet}}
\def\Thor{\textsc{Thor}}
\def\rnsid{\textsc{rns}}

\title
[Runaway Instability]
{The runaway instability in general relativistic accretion disks}

\author[O. Korobkin, et al.]
{O.~Korobkin$^{1,2}$\thanks{E-mail: oleg.korobkin@astro.su.se},
 E.~Abdikamalov$^3$\thanks{E-mail: abdik@tapir.caltech.edu},
 N.~Stergioulas$^4$,
 E.~Schnetter$^{5,6,7}$, 
 B.~Zink$^8$,
 \vspace{0.25cm}\\ 
 {\LARGE\rm S.~Rosswog$^2$, and  C.~D.~Ott$^3$}
 \vspace{0.25cm}\\
 $^1$School of Engineering and Science, Jacobs University Bremen, Germany\\
 $^2$Astronomy and Oskar Klein Centre, Stockholm University, Sweden\\
 $^3$TAPIR, California Institute of Technology, USA\\
 $^4$Department of Physics, Aristotle University of Thessaloniki, Greece\\
 $^5$Perimeter Institute for Theoretical Physics, Canada\\
 $^6$Department of Physics, University of Guelph, Canada\\
 $^7$Center for Computation \& Technology, Louisiana State University, USA\\
 $^8$Theoretical Astrophysics, University of T\"ubingen, Germany}
\date{\today}

\begin{document}
\maketitle
\label{firstpage}
\begin{abstract}
  When an accretion disk falls prey to the runaway instability, a large
  portion of its mass is devoured by the black hole within a few dynamical
  times. Despite decades of effort, it is still unclear under what conditions
  such an instability can occur.
  The technically most advanced relativistic simulations to date were
  unable to find a clear sign for the onset of the instability. 
  In this work, we present three-dimensional relativistic hydrodynamics
  simulations of accretion disks around black holes in dynamical spacetime.
  We focus on the configurations that are expected to be particularly prone
  to the development of this instability.
  We demonstrate, for the first time, that the fully self-consistent
  general relativistic evolution does indeed produce a runaway
  instability.   
\end{abstract}

\section{Introduction}
\label{S:Intro}

Thick and massive relativistic accretion disks around black holes
(BHs) are thought to form in extreme core-collapse events of massive
stars~\citep{Woosley93b, MacFadyen99a, SekiguchiShibata10, Ott11,
  Woosley12}, and they are a normal outcome for the coalescense of
neutron star (NS)-NS~\citep[e.g.,][]{Ruffert96a, Rosswog03b,
  ShibataTaniguchi06, OechslinJanka06a, Baiotti08, Liu08a, Kiuchi09a}
and NS-BH binaries~\citep[e.g.,][]{Rosswog05, ShibataUryu06a,
  Shibata09a, Etienne09a, Chawla10, RuffertJanka10a, Foucart10a}. Such
systems are thought to be candidates for the central engines of
gamma-ray bursts~\citep[GRBs,][]{Popham99a, Woosley93b, Piran04a,
  Lee07, Meszaros12}.   

Previous studies of the stability of accretion disks have shown 
that they can be subject to various types of global instabilities in a
number of scenarios~\citep[e.g.,][]{Abramowicz83, PPI, PPII,
  Kojima86a, WTH94, Font02a, Zanotti03a, TMP10, Korobkin11,
  Kiuchi11a}. Instabilities can result in strongly variable and
unstable accretion rates. \cite{Abramowicz83} discovered the so-called
dynamical runaway instability (RI) in thick, self-gravitating
accretion disks around BHs. The RI is similar to the dynamical
instability in close binary systems that occurs when the more massive
binary member overflows its Roche lobe. In this case, the size of the
Roche lobe decreases faster than the size of the binary companion,
which can ultimately lead to the tidal disruption of the companion and
the merger of the binary system. In disks around BHs, a toroidal
surface analogous to the Roche lobe can be identified. A meridional
cut of this surface exhibits a cusp located at the $L_1$ Lagrange
point. If the disk overflows this toroidal Roche surface, then the
mass-transfer through the cusp will push the cusp outwards, making a
larger fraction of the disk matter unstable to accretion. This drives
the cusp out even further, and leads to an exponential growth of the
mass-transfer rate. As a result, most of the disk gets consumed by the
BH within just a few dynamical times.

\cite{Abramowicz83} found that the development of the RI depends
  on a wide range of parameters, such as the disk-to-BH mass ratio 
$M_\mathrm{D} / M_\mathrm{BH}$ and the location of the inner edge of
the disk\footnote{ Here, the term ``disk'' refers to initial
    equilibrium disk configurations. Therefore, the concept of the
    inner edge for such systems is well defined. Note, however, that
    disk tend to spread on a viscous timescale.}. However, their
investigation was based on several approximations and simplifications:
they used a polytropic equation of state (EOS) for the disk
  material, a pseudo-Newtonian potential to model gravity of
BH~\citep{Paczynsky80}, a disk with constant specific angular
momentum, and an approximate treatment of the self-gravity of the
disk. Subsequent works with more refined approximations found
indications of a stabilizing effect due to BH
rotation~\citep{Wilson84, Abramowicz98}, while a positive radial
gradient of specific angular momentum was suggested to strongly
suppress the instability~\citep{Abramowicz98, Daigne97a, Font02b,
  DaigneFont04}. Moreover, studies using a Newtonian pseudopotential
for the gravity of the BH~\citep{Khanna92, Masuda98a} and relativistic
calculations in a fixed spacetime background~\citep{Nishida96, 
  Font02a} found that the self-gravity of the disk aggravates the
instability. Nevertheless, the distance between the inner edge of the
disk and the location of the cusp is probably the most crucial
parameter for the development of the RI: for as long as it is too
large, the RI is unlikely to occur.  

Recently, \cite{Montero10} performed the first fully general
relativistic simulations of thick accretion disks around BHs in
axisymmetry for a few dynamical times. For the particular models they
studied, they found no signature of a RI during the simulated
time. However, the inner surface of their disk models was located away
from the Roche surfaces (P.~Montero, private communication), so that
the instability might not have had sufficiently favourable conditions
or sufficient time to develop within the time period of disk evolution
that was considered. Therefore, their results do not rule out the
existence of the RI in disk models where the inner edge is located
closer to the cusp.  

In \cite{Korobkin11}, some of us have analyzed the stability
of slender and moderately slender accretion disks around BHs with $
M_\mathrm{D} / M_\mathrm{BH}$ in the range of (0.11, 0.24) using
three-dimensional (3D) numerical simulations in full general
relativity (GR\@). Although we did observe the development of several
non-axisymmetric instabilities in our models, we found no traces of
the RI\@. This result is, perhaps, not surprising since the inner
radii of our disk models were located significantly away from the
Roche surface (see \cite{Korobkin11} for more details).   

In a similar study, \cite{Kiuchi11a} performed simulations of
self-gravitating disks around BHs in full GR with the aim of
obtaining the gravitational wave signal from the non-axisymmetric
Papaloizou-Pringle instability. They considered four disk models with
constant and non-constant specific angular momentum and disk-to-BH
mass ratios of $0.06$ and $0.10$. No runaway instability was
observed in their simulations, but no information was given
regarding the relative location of the inner edge of the disk and
the Roche surface. Therefore, it is difficult to judge if their
models were sufficiently susceptible to develop the RI.

In order to understand whether the RI can occur at all in the most
general case, one first has to expore whether it can develop in configurations
that are \emph{particularly prone} to the instability.
Such systems contain disks that exactly fill their Roche lobes,
have significant fractions of the BH masses, constant specific angular
momentum profiles and non-rotating BHs\footnote{
BH+disk systems with slowly rotating BHs are unlikely to be
drastically different from non-rotating BH cases.
Nevertheless, the latter are expected to be more susceptible to the RI,
that is why they are in the focus of our work.
}.
It is left to future studies to explore under which circumstances such
configurations would form in Nature. Our aim here is to explore whether the RI can occur
\emph{in principle} if the fully dynamical general-relativistic effects are taken into
account properly.
Our 3D general relativistic simulations indeed confirm that the RI occurs in
these cases.

This paper is organized as follows. In
Section~\ref{S:NumericalMethods}, we describe the numerical methods
used in our simulations. In Section~\ref{S:DiskModels}, we describe
the initial disk model. In Section~\ref{S:Results}, we present our
results, and in Section~\ref{S:Conclusion} present our
conclusions for future work.
Unless otherwise noted, throughout the paper we measure distances in
the units of the Schwarzschild radius $r_g = 2GM_{\rm BH}(0)/c^2$ of
the black hole at $t=0$, while the rest of the quantities are measured
in cgs units.  

\section{Numerical methods}
\label{S:NumericalMethods}

The numerical time-evolution in our study is performed with the
GR hydrodynamics code \Thor{}~\citep{Zink08} and the spacetime evolution
code \Quilt{}~\citep{Pazos07}. These two codes are based on and
communicate with each other via the open-source \Cactus{} computational
infrastructure~\citep{Goodale02a, Cactus_DBLP}, and use the \Carpet{}
mesh refinement and multiblock driver~\citep{Schnetter04a,
  Schnetter:2006pg}. Initial models of self-gravitating disks around BHs
are constructed using an improved version of the \rnsid{}
code~\citep{Stergioulas11}. 
We use a multiblock grid with an adapted resolution which is similar to the 
one used in~\cite{Korobkin11}.
The grid resolution is $\Delta_{\rm min}\sim0.022\,r_g$ near the BH,
$\Delta_{\rm max}\sim0.2\,r_g$ at the disk density maximum, and 
$\Delta_{\rm out}\sim5.0\,r_g$ at the outer boundary.
For more details of the numerical methods
as well as the grid structure used in our simulations, we refer the reader
to~\cite{Korobkin11}.

\section{Initial disk model}
\label{S:DiskModels}

  \begin{table}
  \begin{center}
  {\footnotesize
  \begin{tabular}{l|c|c}
  \hline
  Model                                                      &   A    &   B    \\
  \hline
  Specific angular momentum $\ell \ [M_{BH}]$                & 3.91   & 3.84   \\
  Polytropic constant $K \ [10^{14}\cm^3\gr^{-1/3}\sec^{-2}]$& 5.35   & 5.72   \\
  Maximum density $\rho_\mathrm{max} \ [10^{12}\gr/\cm^3]$   & 1.63   & 1.33  \\
  Disk-to-BH mass ratio $M_D/M_{BH}$                         & 0.2097 & 0.1628 \\
  Kinetic to potential energy $T/|U|$                        & 0.2666 & 0.2469 \\
  Inner radius $r_{\rm in} [r_g]$                            & 2.082  & 2.127  \\
  Location of the cusp $r_L [r_g]$                           & 2.024  & 2.148  \\
  Outer radius $r_{\rm out} [r_g]$                           & 12.08  & 12.37  \\
  Central radius $r_\mathrm{max} [r_g]$                      & 4.687  & 4.342  \\
  Orbital frequency at $r_\mathrm{max}$, $\Omega_\mathrm{max} \ [\sec^{-1}]$  & 1414   & 1581   \\
  Orbital period at $r_\mathrm{max}$, $P_\mathrm{max} \ [\sec]$   & $4.44\cdot10^{-3}$  &  $3.96\cdot10^{-3}$ \\
  Potential gap $\Delta W:=W_{\rm in}-W_L$                   & $-1.2\cdot10^{-4}$  & $+0.01$  \\
  \hline
  \end{tabular}
  }
  \end{center}
  \caption{Physical parameters of the initial disk models used. Notice
    that the specific angular momentum and $\Delta W$ are given in
    units of $c=G=1$.}   
  \label{T:modelparams}
  \end{table}

\begin{figure}
  \begin{center}
    \includegraphics[width=0.49\textwidth]{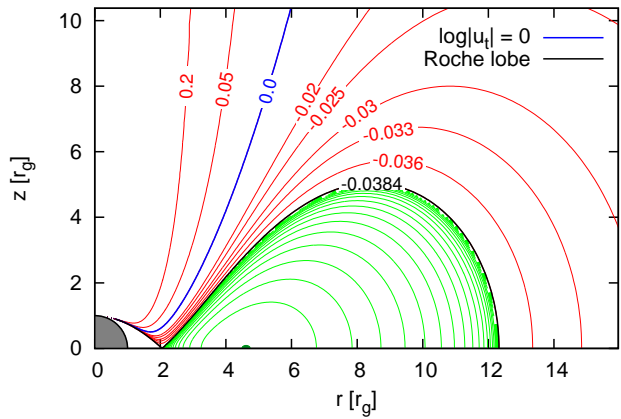} 
    \caption{The structure of the initial disk model A\@.
    Red and green lines show the contours of the effective potential
    $W=\log|u_t|$ and of the disk pressure, respectively.
    Here, $u_t$ is the covariant $t$-component of the $4$-velocity
    of a test particle on a circular orbit with the given specific
    angular momentum for model A.
    The values of $W$ are given in geometrized units $G = c = 1$. 
    Contours of the pressure are equally spaced by $0.5$ in decimal 
    logarithmic scale, 
    starting from the maximum pressure (green dot on the $r$-axis).
    The blue contour corresponds to $W=0$, and the black contour marks 
    the location of the toroidal Roche lobe with the cusp.
    The disk fills the Roche lobe almost up to the cusp.
    The gray area marks the location of the BH event horizon.}
    \label{F:id}
  \end{center}
\end{figure}

In this study, we focus on two initial disk models, denoted A and B\@.
Table~\ref{T:modelparams} lists the parameters of these models, while
Fig.~\ref{F:id} illustrates the structure of initial model A\@. 
We construct our disks using a polytropic EOS $p=K\rho^{\Gamma}$ with
$\Gamma=4/3$, while the time-evolution is performed with the
two-parameter ideal gas EOS (with the same $\Gamma$), which allows for
shock heating. We do not include additional physics such as magnetic
fields, nuclear or neutrino processes. Therefore, the results can be
rescaled using one free parameter such as the initial BH mass. 
In particular, the cgs parameters listed in Table~\ref{T:modelparams}
assume $\Mbh=5\;\MSun$.  
The disk-to-BH mass ratio $\Mdisk/\Mbh$ is $\approx0.21$ for model A
and $\approx0.16$ for model B\@. Both disks have a constant specific
angular momentum profile. 
The black contour in Fig.~\ref{F:id} corresponds to the Roche surface,
while  the green ones show the contours of constant pressure. The disk
in model A fills its Roche lobe almost entirely, with a remaining gap
in the effective potential between the inner edge of the disk and the
cusp of only $\Delta W=-1.2\times10^{-4}$. By contrast, model B is
constructed to slightly overfill its Roche lobe by the value $\Delta
W=0.01$. This is done in order to induce the onset of the RI right
from the very beginning of the numerical evolution.   

\section{Results}
\label{S:Results}

\begin{figure}
  \begin{center}
    \includegraphics[width=0.49\textwidth]{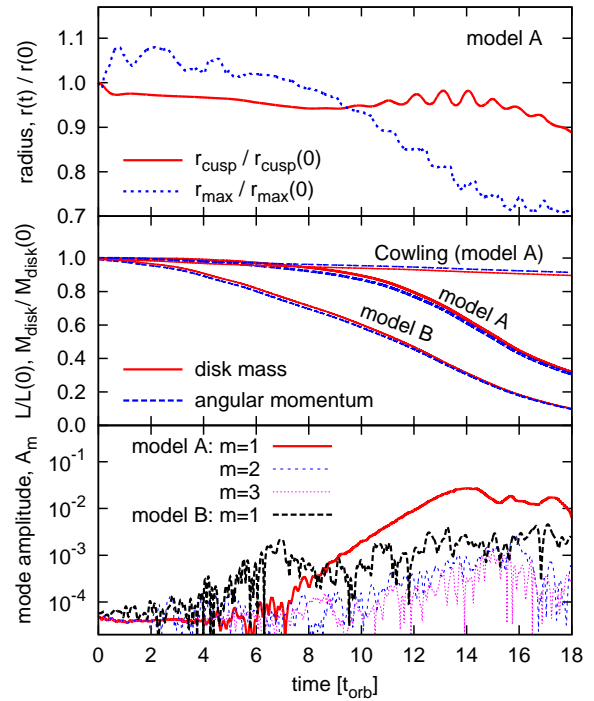} 
    \caption{Evolution of the disk.
    {\it Top}: radial location of the cusp and the disk density maximum, normalized
    to their initial values.
    {\it Centre}: evolution of the disk rest mass and angular momentum
    for models~A and~B in full GR as well as for model A in the
    Cowling approximation.
    For clarity, the deviations from $1$ of the mass and angular momentum
    in Cowling are multiplied by a factor of 10.
    {\it Bottom}: Normalized amplitudes of non-axisymmetric $m=1,2,3$
    deformations in model~A and $m=1$ deformation in model~B\@. 
    }
    \label{F:diskBHmassam}
  \end{center}
\end{figure}

Models A and B both develop the runaway instability. Therefore, in the
following, we concentrate on the results for model A, while model B
will be discussed later in this section.
The top panel of Fig.~\ref{F:diskBHmassam} shows the time evolution
of the cusp radius and the radial coordinate of the location of the
maximum disk density $r_\mathrm{max}$. Due to initial metric
perturbations induced by matching the initial data to a vacuum BH
metric near the horizon \citep[cf.\ discussion in][]{Korobkin11}, 
the BH mass and cusp radius settle to a new, $\sim 3 \%$~smaller
value within $t \lesssim 0.5 t_\mathrm{orb}$. The smaller
gravitational pull of the BH leads to a rapid increase of
$r_\mathrm{max}$ by $\sim 9 \%$ within the same time interval. Since
the cusp is now located at a smaller radius, the disk is less likely
to become subject to the RI\@. However, the initial metric perturbations
induce oscillations in the disk, which are particularly evident in the
evolution of $r_\mathrm{max}$ at $t \lesssim 6 t_\mathrm{orb}$. These
oscillations lead to occasional crossing of the Roche lobe by the
disk, resulting in a steady and slow accretion of the disk
material onto the BH at $t \lesssim 9 t_\mathrm{orb}$. This is visible
in the evolution of the disk mass and angular momentum shown in the
center panel of Fig.~\ref{F:diskBHmassam}.

\begin{figure*}
  \begin{center}
    \includegraphics[width=0.95\textwidth]{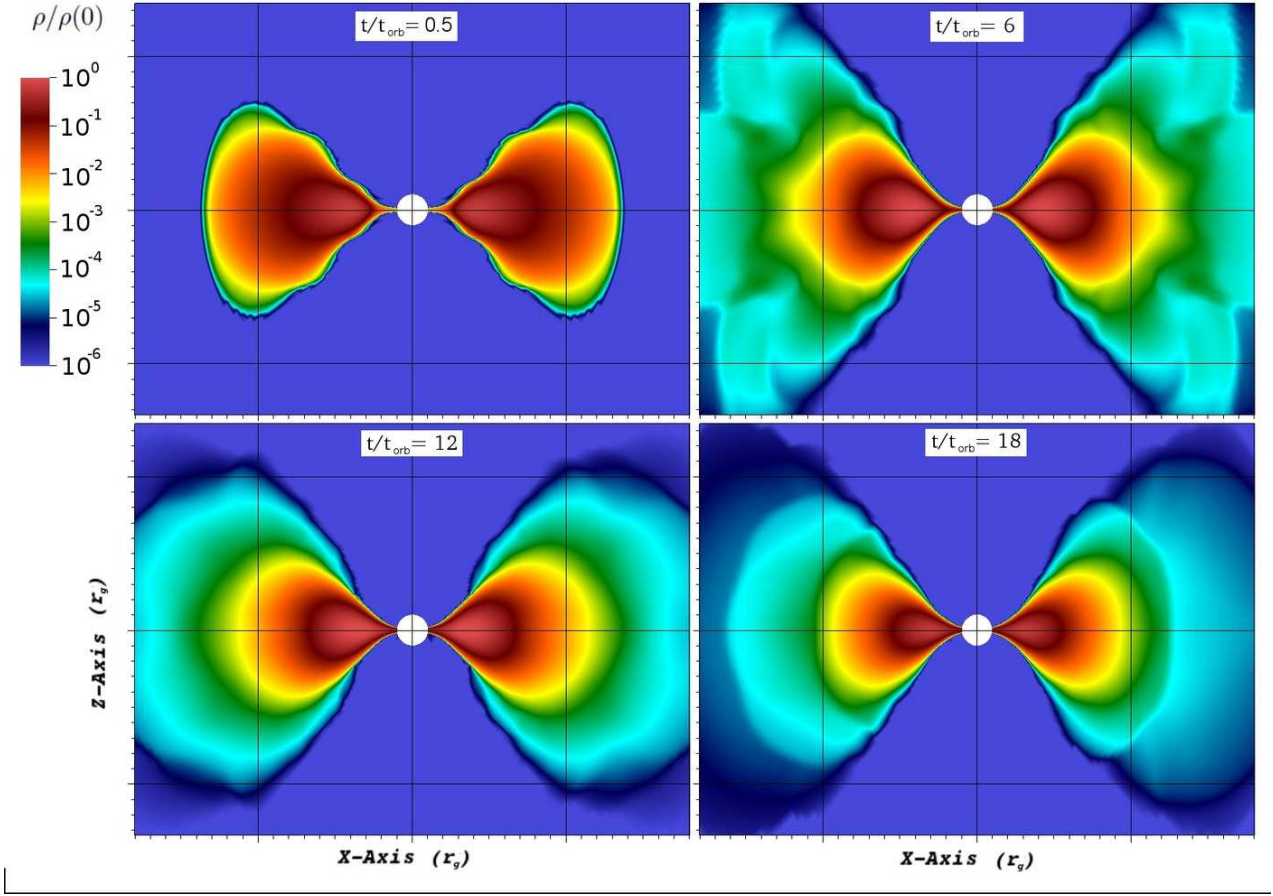} 
    \caption{Colourmaps of the disk density in the vertical plane at
      different times in evolution of model A. Here, $t_{\rm orb}$ is
      the orbital period at the maximum density radius $r_{\rm max}$.}   
    \label{F:colorframes}
  \end{center}
\end{figure*}

The radius $r_\mathrm{max}$ gradually decreases due to this slow
accretion up to the 
time when the inner disk radius becomes as small as the
cusp radius. This occurs at $t \sim 9 t_\mathrm{orb}$. At that point, 
the cusp radius starts increasing, leading to an acceleration of the
accretion. By the end of our simulation ($t = 18 t_\mathrm{orb}$),
$\sim 75 \%$ of the initial disk mass has been accreted onto the
BH\@. Such an accelerated accretion due to dynamical migration of the
cusp towards the disk during which most of the disk material is
swallowed by the BH is exactly the defining property of the runaway 
instability. Thus, our
simulations show that the RI can indeed occur, at least for the
  most susceptible models, which are considered here.

It is interesting to note that the development of the axisymmetric
runaway instability triggers non-axisymmetric deformations of the
disk. The bottom panel of Fig.~\ref{F:diskBHmassam} shows the normalized
amplitude of non-axisymmetric $m=1,2,3$ deformations (see
\cite{Korobkin11} for the exact definition of the amplitude).  
The $m=1$ deformation grows exponentially starting at $\sim 7
t_\mathrm{orb}$ until $\sim 14 t_\mathrm{orb}$, reaches its peak value
of $\sim0.027$, at which point it saturates. This deformation 
develops due to the so-called Papaloizou and Pringle instability
(PPI)~\citep{PPI}, enhanced by an eccentric motion of the BH~\cite[as
  described in][]{Korobkin11}. The deformations corresponding to the
other values of $m$ do not show a strong growth and remain below $\sim
10^{-3}$. Since both the PPI and RI develop roughly at the same time
in our simulations, they could, in principle, interact nonlinearly,
when sufficiently large amplitudes are reached. In particular, the
PPI-induced deformations might be responsible for the apparent
saturation of the RI towards the end of our simulations. Such
deformations redistribute angular momentum outwards, which, in turn,
inhibits the development of the RI~\citep{DaigneFont04} (while another
factor responsible for the eventual saturation of the RI is, of course,
the depletion of the disk mass). Nevertheless, the rather moderate
amplitude of the $m=1$ mode of $\sim0.027$ is unlikely to drastically
affect the evolution of the RI\@, especially during the early
evolution. 
Neither can it be held responsible for the onset of the RI itself,
since the accretion stream overflowing the cusp and triggering the RI is
only slightly non-axisymmetric.
A more detailed analysis of the non-linear interaction between
the PPI and the RI is required to clarify its role. 

Figure~\ref{F:colorframes} shows four snapshots of the disk density in
the meridional plane, for model A, corresponding to the times
$t/t_{\rm orb}=0.5$, $6$, $12$ and $18$. The snapshot at
$t=0.5\;t_{\rm orb}$ shows the disk in a perturbed state after the
passage of the initial metric perturbation. As the disk moves away
from the BH, several surface waves can be seen propagating to the
outer side of the disk in an axisymmetric manner. As noted above, such
waves have sufficiently high amplitude to overfill the Roche lobe and
support a small amount of accretion. The next snapshot at $t=6\;t_{\rm
  orb}$ shows a wider accretion stream and an extended structure of
the outer parts of the disk, caused by heating by the shocks formed
during the radial disk oscillations. Such shock heating efficiently
damps out the radial disk oscillation and causes the disk to heat up
and expand, similarly to what was observed in~\cite{Korobkin11}. 
This thermal expansion is confined to the outer regions of the disk.
Indeed, if the expansion significantly affected the inner regions,
the disk would overflow its Roche lobe and additional streams from the
upper and lower faces would emerge. Since this is not observed 
on~\ref{F:colorframes}, we conclude that the inner side of the disk is 
not affected. The
last two snapshots show the disk during the rapid accretion phase at
$t=12\;t_{\rm orb}$ and $t=18\;t_{\rm orb}$. By $t=18\;t_{\rm orb}$,
the accretion stream is very wide, but the high-density parts have
moved closer to the black hole and occupy a smaller volume, indicating
a smaller disk mass. 

To secure our findings against the possibility of an error in
the treatment of hydrodynamics, we have complemented the fully
dynamical GR simulation of model A with one in the Cowling
approximation (i.e., on a fixed metric background). Such a model
should undergo at most a slow and steady accretion, in a way similar
to the early ($t \lesssim 6 t_\mathrm{orb}$) evolution of the fully
dynamical model. There should be no unstable growth of accretion on a
fixed metric background. As we can see from the time evolution of the
disk mass and angular momentum shown in the center panel of
Fig.~\ref{F:diskBHmassam}, this is indeed the case.   

Interestingly, model A does not develop the PPI in our Cowling
simulation, where amplitudes of non-axisymmetric deformations remain 
below $\sim10^{-5}$ throughout the evolution. Since the PPI in Cowling
approximation relies on the existence of the non-accreting inner edge of the
disk~\citep{Blaes87a,Hawley91:BHTori3D}, the absence of PPI is
probably caused by the accretion at a small rate through the
cusp, which can completely freeze the development of PPI modes in 
Cowling approximation~\citep{Blaes87a,Hawley91:BHTori3D}. 
This is different in the case of a dynamical metric, where the PPI can 
be enhanced by the
motion of the BH~\citep{Korobkin11}. For this reason, accretion fails
to suppress the PPI in model A with a dynamical metric (this is also
observed in Fig.~1 of \cite{Kiuchi11a} for the models with constant
specific angular momentum).

Compared to model A, in model B the disk overflows its Roche
lobe by a small amount $\Delta W=0.01$. In this case, the 
exponential accretion should commence immediately and continue until a
significant fraction of the disk is accreted onto the BH within a few
dynamical times. This is what we observe in our simulation, as can be
seen from the evolution of the disk mass and angular momentum shown in
the center panel of Fig.~\ref{F:diskBHmassam}. Notice, however, that
in this case the development of the instability happens more slowly,
because of the lower disk-to-BH mass ratio. The amplitude of
non-axisymmetric deformations is also lower. In particular, the $m=1$
mode exhibits slow and irregular growth and stays below $\sim 10^{-3}$
throughout the simulation (cf. the center panel of
Fig.~\ref{F:diskBHmassam}). In the absence of any significant
non-axisymmetic deformation, the accretion of a substantial fraction
of the disk mass onto the BH within a few dynamical times can only be
attributed to the development of the RI.

As mentioned in Section~\ref{S:Intro}, \cite{Montero10} presented
axisymmetric simulations of two disk models around BHs with a constant
specific angular momentum. Their simulations were also fully general
relativistic with dynamical spacetime evolution, similar to our
simulations. 
The absence of the RI in \cite{Montero10} must be attributed to 
differences in the initial disk models (compared to our configurations)
and specifically to the fact that their initial models did
not exactly fill the Roche lobe. Because of this, the instability is
less likely to occur within the limited simulation time. The same
argument may explain the absence of the RI in~\cite{Kiuchi11a}. If
confirmed, this would further underline the importance of the distance
between the cusp and the inner disk surface for the development of 
RI.~\footnote{The RI was also not observed in full GR simulations of the
accretion disks formed in the aftermath of the merger of NSs with their
binary compact companions~\cite[e.g.][]{Rezzolla10a}. We believe the same
rationale applies also here.}

\section{Conclusion}
\label{S:Conclusion}

In this study, we have, for the first time, demonstrated that the runaway
instability in self-gravitating accretion disks does indeed occur in
fully dynamical general relativistic evolutions. We have selected two
models that are particularly prone to the development of such an
instability. Our models have an appreciable disk-to-BH mass ratio
($0.21$ for model A and $0.16$ for model B, see
Table~\ref{T:modelparams}), a constant profile of specific angular
momentum and a non-rotating BH\@. Moreover, and perhaps most
importantly for the development of the runaway instability, our disk
model A almost exactly fills its Roche lobe, while model B slightly
overfills it.    

Our simulations show that both models develop the runaway instability,
exhibiting unstable accretion of the disk matter onto the BH within
just a few dynamical times. More than half of the disk mass is
absorbed by the BH by the end of our simulations, which were terminated
only because we ran out of available computing time.  

Our results demonstrate that the runaway instability does indeed occur, at
least in the models considered here. Future research will need to
investigate how this depends on the parameters of the initial disk
models, such as the disk-to-BH mass ratio, the gradient of the specific
angular momentum of the disk, and the BH spin, in order to establish
the astrophysical significance of the instability. 

Complementary animation illustrating the development of the runaway 
instability can be downloaded from: 
\url{http://compact-merger.astro.su.se/movies.html#accrd}.

\section{Acknowledgements}
\label{S:Acknowl}

We acknowledge stimulating discussions with P.~Diener, P.~Montero, C.~Reisswig,
M.~Scheel, B. Szil\'{a}gyi, and J.~Tohline. This work is supported by
the National Science Foundation under grant numbers AST-1212170, 
PHY-1151197, PHY-1212460, and OCI-0905046, by the German Research
Foundation grant DFG RO-3399, AOBJ-584282, and by the Sherman 
Fairchild and Alfred P. Sloan Foundation. NS acknowledges support by
an Excellence Grant of the research committee of the Aristotle
University of Thessaloniki. Supercomputing simulations for this
article were performed on the Compute Canada SHARCNET cluster ``Orca''
(project CFZ-411-AA), Caltech compute cluster ``Zwicky'' (NSF MRI award
No.\ PHY-0960291), on the NSF XSEDE network under grant TG-PHY100033,
on machines of the Louisiana Optical Network Initiative under grant
loni\_numrel07, and at the National Energy Research Scientific
Computing Center (NERSC), which is supported by the Office of Science
of the US Department of Energy under contract DE-AC03-76SF00098.

\bibliographystyle{mn2e.bst}
\bsp

\end{document}